\documentclass[aps,prd,reprint,superscriptaddress,amsmath,amssymb,nofootinbib,longbibliography]{revtex4-2}

\usepackage{graphicx}
\usepackage{bm}
\usepackage{booktabs}
\usepackage{microtype}
\usepackage[colorlinks=true,allcolors=blue]{hyperref}

\newcommand{\dd}{\mathrm{d}}

\newcommand{\Lie}{\mathcal{L}}
\newcommand{\Kfi}{\mathcal{K}_{fi}}
\newcommand{\bbra}[2]{\!\left\langle\!\left\langle #1,#2\right\rangle\!\right\rangle\!}
\newcommand{\MaxKernelCorrection}{21.67\%}
\newcommand{\MaxDepletionChange}{13.7\%}

\newcommand{\TidalDomainCount}{38}
\newcommand{\GeometryIncrementRange}{0.031--0.207\%}
\newcommand{\RadialShareRange}{78.7--82.5\%}
\newcommand{\NormalizationShareRange}{13.3--16.2\%}
\newcommand{\MultipoleRepresentativeShift}{0.066\%}
\newcommand{\IndependentKernelDifference}{3.24\times10^{-6}}

\begin{document}

\title{Relativistic Tidal Transitions of Saturated Kerr Boson Clouds}

\author{Yi-kun Li}
\affiliation{State Key Laboratory of Radio Astronomy and Technology,
Xinjiang Astronomical Observatory, CAS, 150 Science 1-Street,
Urumqi 830011, China}
\affiliation{College of Astronomy and Space Science,
University of Chinese Academy of Sciences, No.1 Yanqihu East Road,
Beijing 101408, China}

\author{Lang Cui}
\email{cuilang@xao.ac.cn}
\affiliation{State Key Laboratory of Radio Astronomy and Technology,
Xinjiang Astronomical Observatory, CAS, 150 Science 1-Street,
Urumqi 830011, China}
\affiliation{College of Astronomy and Space Science,
University of Chinese Academy of Sciences, No.1 Yanqihu East Road,
Beijing 101408, China}
\affiliation{Xinjiang Key Laboratory of Radio Astrophysics,
150 Science 1-Street, Urumqi, Xinjiang, 830011, China}

\date{July 16, 2026}

\begin{abstract}
Rotating black holes can bind ultralight bosons in macroscopic clouds whose
level structure is swept by the tidal field of a binary companion.  Existing
estimates of the resulting resonant transitions have largely used
hydrogenic wave functions, even where the cloud grows most efficiently and
the Kerr geometry appreciably deforms its quasibound states.  We compute
tidal transition matrix elements on the numerically determined saturation
branches of the $|211\rangle$, $|322\rangle$, and $|433\rangle$ clouds.  The
calculation combines Kerr quasibound modes, their relativistic bilinear
normalization, and an adiabatic quadrupolar perturbation of the Kerr metric.
Across five $\Delta m=-2$ channels, the relativistic matrix element differs
from its hydrogenic value by as much as $\MaxKernelCorrection$.  Corrections
above $10\%$ persist when $99\%$ of the cloud lies within $7\%$ of the
orbital separation.  A component-resolved comparison attributes
$\RadialShareRange$ of the logarithmic change to the radial mode profile and
$\NormalizationShareRange$ to the bilinear normalization; the relativistic
tidal geometry supplies a smaller additional correction.  Exact Newtonian
multipoles and an independent complex-contour construction confirm the
quadrupole matrix elements.  For resonances with intermediate
Landau--Zener adiabaticity, the corrected matrix elements change the
predicted cloud depletion by up to $\MaxDepletionChange$.  Relativistic cloud
structure is therefore quantitatively important for binary histories that
cross saturated-cloud resonances.
\end{abstract}

\maketitle

\section{Introduction}
\label{sec:introduction}

Rotational superradiance turns a Kerr black hole into a spectroscopic probe
of light bosonic fields.  The amplification mechanism was recognized in
rotating media and black-hole scattering
\cite{Zeldovich1971,PressTeukolsky1972,Teukolsky:1972my,Teukolsky:1973ha,
Teukolsky:1974yv},
while the massive Klein--Gordon instability established that a field's mass
can provide the confining barrier required for exponential growth
\cite{Detweiler:1980uk,Dolan:2007mj}.  For a boson of mass $\mu$ around a
black hole of mass $M$, the coupling $\alpha=M\mu$ plays the role of a
gravitational fine-structure constant.  At small $\alpha$, the bound states
resemble hydrogenic orbitals.  Superradiant levels acquire enormous
occupation numbers and form a coherent ``gravitational atom'' whose size is
much larger than the horizon \cite{Arvanitaki:2009fg,Arvanitaki:2010sy,
Brito:2015oca}.

The atomic analogy organizes the cloud by integers $(n,l,m)$, but the Kerr
problem also carries a distinctive hierarchy.  The leading binding energy is
set by $\alpha^2$, relativistic fine structure separates states with different
$l$, and frame dragging produces an $m$-dependent hyperfine splitting.
Meanwhile, the superradiant condition $\omega_R<m\Omega_H$ selects positive
$m$ states and transfers spin into them until equality is reached.  This
equality fixes the saturation spin of a mature cloud.  The mode's radial
extent scales approximately as $n^2M/\alpha^2$ in the weak-field regime, so a
cloud can be spatially dilute while its inner tail and frequency already
carry measurable relativistic corrections.  The coexistence of a large
orbital size and a strong-field boundary condition makes the transition
amplitude sensitive to relativistic information beyond hydrogenic power
counting.

This phenomenon connects black-hole dynamics to particle physics and
gravitational-wave astronomy.  Measurements of black-hole masses and spins
can exclude ranges of boson masses \cite{Brito:2014wla,Cardoso:2018tly,
Baryakhtar:2017ngi}.  Annihilations, level transitions, and nonlinear cloud
dynamics can also produce long-lived or transient gravitational radiation
\cite{Brito:2017wnc,Brito:2017zvb,Hannuksela:2018izj,Isi:2018pzk,
Siemonsen:2019ebd,Siemonsen:2022yyf,Tsukada:2020lgt,Peng:2025zca}.
Numerical evolutions
have clarified the development and collapse of bosonic clouds
\cite{Yoshino:2012kn,Witek:2012tr,Dolan:2012yt}, and analogous instabilities
of vector fields enlarge the range of growth rates and signals
\cite{Pani:2012vp,Dolan:2018dqv,Percival:2020skc}.  Self-interactions can
alter the late-time occupation while leaving the linear level structure as
the natural starting point \cite{Baryakhtar:2020gao}.

A binary companion adds a slowly rotating tidal perturbation to this atom.
As the orbital frequency increases, a Fourier component of the tide becomes
stationary relative to two cloud modes and drives a resonant transition.
The associated transfer of angular momentum can deplete the cloud, ionize
it, or feed angular momentum back to the orbit, in some cases supporting a
floating orbit \cite{Zhang:2018kib,Baumann:2018vus,Baumann:2019ztm,
Baumann:2021fkf,Baumann:2022pkl}.  This picture has been developed from
hydrogenic spectra through relativistic level splittings
\cite{Baumann:2019eav,Berti:2019wnn}, numerical studies of tidal disruption
\cite{Cardoso:2020hca}, and dissipative multilevel dynamics
\cite{Tong:2022bbl,Takahashi:2023}.  More recent analyses follow resonant
histories, wide transitions, and coupled cloud-orbit evolution
\cite{Tomaselli:2023ysb,Tomaselli:2024bdd,Tomaselli:2025wide,
Boskovic:2025trails}.  Together they show that transition strengths are a
central microscopic input to the macroscopic history of a boson cloud.

The resonance is especially sensitive to the transition matrix element.
For a level separation $\Delta\omega$, the selected tidal harmonic crosses
resonance when $\Delta\omega=\Delta m\,\Omega$.  During a slow crossing, the
square of the matrix element competes with the orbital sweep rate in the
Landau--Zener exponent.  A modest amplitude error is therefore doubled at
the level of this exponent and can move an intermediate crossing between
substantial survival and substantial depletion.  The same coupling sets the
width of dissipative resonances and the torque exchanged with the orbit.
Accuracy at the ten-percent level is consequently relevant even before a
complete waveform model is attempted.  It can change which level is
populated after the crossing and thus which later resonances appear in the
binary history.

Two physical scales make this precision problem nonuniform.  Fine and
hyperfine transitions occur at frequencies parametrically below the boson's
orbital frequency inside the cloud, placing the companion far from the black
hole.  This favors a multipolar, adiabatic treatment of the external field.
At the same time, the initial cloud is selected by superradiant saturation,
and its quasibound boundary condition reaches the horizon.  The external
perturbation can therefore be weak and slowly varying while the states on
which it acts remain genuinely relativistic.  The accuracy of the tidal field
and the cloud modes must therefore be controlled independently.

Hydrogenic matrix elements become least secure precisely where several
observationally interesting clouds grow rapidly.  Kerr frame dragging changes
the radial and angular quasibound functions, the real frequencies determine
the resonance location, and a relativistic bilinear form sets the mode
normalization.
Relativistic perturbation theory has recently made these effects calculable
for black-hole boson clouds \cite{Cannizzaro:2023jle}.  Conserved currents
provide the required mode orthogonality and analytic continuation
\cite{Green:2022htq}, and the same framework has been extended to forced
quasinormal-mode excitation \cite{Cannizzaro:2025qnm}.  In parallel, local
tidal solutions for spinning black holes have advanced from slowly rotating
metrics \cite{YunesGonzalez2006,Poisson:2014gka,Landry:2015zfa} to
adiabatic Kerr geometries reconstructed in radiation gauge
\cite{Katagiri:2026,Cocco:2026}.  These developments permit the bound-state
and tidal sides of the transition problem to be treated within one covariant
projection.

The relativistic projection introduces conceptual differences in addition
to numerical ones.  Kerr quasibound modes have complex frequencies and
radiative boundary conditions.  These properties call for a conserved
bilinear form that uses analytic continuation through the near-horizon
region.  A tidal matrix element must use the same continuation to preserve
mode orthogonality.  Its
complex phase depends on the arbitrary phases of the two modes, whereas its
magnitude controls an isolated two-level transition.  These facts motivate
both a covariant formulation of the perturbing operator and an independent
contour calculation of the final matrix element.

The missing element is a comparative spectrum of relativistic
quasibound-to-quasibound matrix elements on the physically selected
saturation branches.  Existing relativistic examples establish that
wave-function effects can be appreciable.  These results motivate a
comparative map of the principal $|211\rangle$, $|322\rangle$, and
$|433\rangle$ clouds across their multiplets.  Such a map must separate
changes due to the Kerr cloud from those due to the tidal geometry, because
these ingredients have different radial support and different implications
for improving binary models.

This paper supplies that spectrum for five corotating $\Delta m=-2$
transitions.  The main results are threefold.  First, exact Kerr modes reduce
the transition amplitude by up to $\MaxKernelCorrection$ relative to the
hydrogenic quadrupole prediction, and corrections above $10\%$ remain in a
conservative region where the cloud occupies a small fraction of the binary
separation.  Second, controlled component replacements show that the radial
mode profile produces most of the change, followed by the relativistic
bilinear normalization; the Kerr tidal geometry changes the amplitude by
only $\GeometryIncrementRange$ over the same sample.  Third, exact
Newtonian multipoles, independent contour-normalized modes, and a
Schwarzschild comparison establish that the dominant effect is stable under
changes in the tidal expansion and radial construction.  When propagated
through a prescribed inspiral, the matrix-element correction changes the
Landau--Zener depletion by as much as $\MaxDepletionChange$.

The presentation is organized as follows.  Section~\ref{sec:clouds}
introduces saturated Kerr clouds and the selected resonances.
Section~\ref{sec:matrix} develops the relativistic projection.
Section~\ref{sec:spectrum} presents the transition spectrum and its physical
origin.  Section~\ref{sec:robustness} examines spatial validity, higher
multipoles, independent mode construction, and binary consequences.
Section~\ref{sec:discussion} discusses the implications.  Appendices
\ref{app:qbs}--\ref{app:lz} collect the field equations, normalization,
gauge argument, selection rules, numerical comparisons, and inspiral
formulae.

\section{Saturated Kerr clouds and tidal resonances}
\label{sec:clouds}

We use units $G=c=\hbar=1$ and metric signature $(-,+,+,+)$.  The black-hole
mass sets the unit of length and time, and $\chi=a/M$ denotes its
dimensionless spin.  A complex representation of a real massive scalar
field separates as
\begin{equation}
 \begin{aligned}
 \Phi_{nlm}&=e^{-i\omega_{nlm}t+im\phi}
 S_{lm}(\theta;c)R_{nlm}(r),\\
 c^2&=a^2(\omega_{nlm}^2-\mu^2).
 \end{aligned}
 \label{eq:mode}
\end{equation}
The radial solution is ingoing at the future horizon and exponentially
decaying at infinity.  These boundary conditions select discrete complex
frequencies.  The real part gives the level energy and the imaginary part
gives growth or absorption.  We determine both from the massive-scalar
continued fraction \cite{Leaver:1985ax,Dolan:2007mj}; the separated
equations and convergence conditions are summarized in
Appendix~\ref{app:qbs}.  Angular eigenvalues are followed continuously along
the scalar spheroidal branch \cite{Berti:2005gp}.

Superradiant extraction ends when the mode co-rotates with the horizon.
Each initial state is consequently evaluated on its own saturation branch,
defined by
\begin{equation}
 \begin{aligned}
 \mathrm{Re}\,\omega_i(\alpha,\chi_{\rm sat})
 &=m_i\Omega_H(\chi_{\rm sat}),\\
 \Omega_H&=\frac{\chi}{2M(1+\sqrt{1-\chi^2})}.
 \end{aligned}
 \label{eq:saturation}
\end{equation}
This construction accounts for the relativistic displacement of the
saturation spin.  It also selects a common physical stage of cloud evolution
for comparing different values of $\alpha$.

\begin{figure*}[t]
 \includegraphics[width=\textwidth]{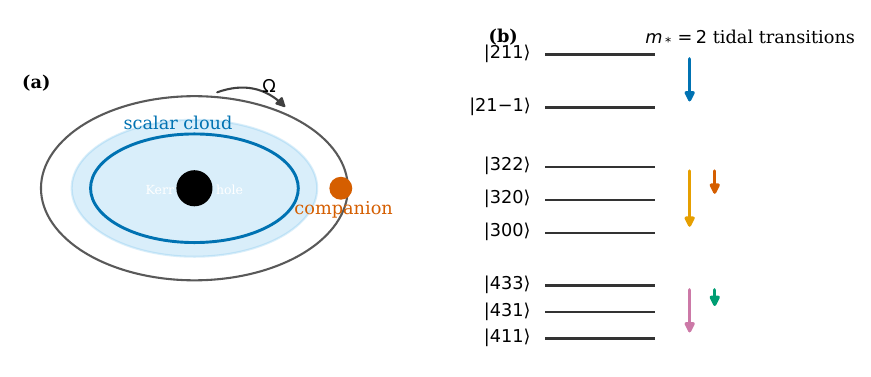}
 \caption{Physical setting and selected transitions.  (a) A saturated Kerr
 black hole is surrounded by a scalar cloud and perturbed by an equatorial
 companion.  The orbital harmonic with $m_*=2$ rotates through the cloud at
 $\Omega$.  (b) The five channels connect initial superradiant states to
 decaying or nonsuperradiant states.  Arrows encode only the level ordering.}
 \label{fig:setup}
\end{figure*}

For a circular equatorial companion, a tidal harmonic with azimuthal number
$m_*=m_i-m_f$ drives a transition when its phase matches the level
splitting,
\begin{equation}
 M\Omega_{\rm res}=
 \frac{\mathrm{Re}[M(\omega_i-\omega_f)]}{m_i-m_f}>0.
 \label{eq:resonance}
\end{equation}
The quadrupolar selection rules require $|l_i-l_f|\le2\le l_i+l_f$,
$l_i+l_f$ even, and $\Delta m=-2$ for the corotating harmonic used here.
They lead to the five channels shown in Fig.~\ref{fig:setup}: the hyperfine
transitions $211\to21{-1}$, $322\to320$, and $433\to431$, together with the
cross-multiplet transitions $322\to300$ and $433\to411$.  The first three
probe different principal levels at fixed $l$, whereas the last two are more
sensitive to cancellations between distinct radial profiles.

\begin{figure*}[t]
 \includegraphics[width=\textwidth]{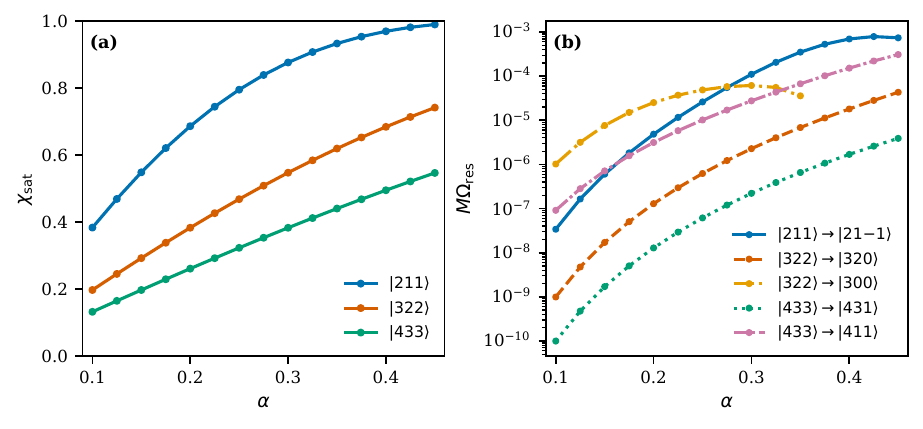}
 \caption{Saturation and resonance scales.  (a) Numerically determined
 saturation spins for the $|211\rangle$, $|322\rangle$, and $|433\rangle$
 clouds.  (b) Resonance frequencies for the five $\Delta m=-2$ transitions.
 Panel (b) shows only prograde resonances with $M\Omega_{\rm res}>0$.}
 \label{fig:saturation}
\end{figure*}

The local tidal description is controlled by three scale comparisons.  For
each participating mode $s=i,f$, we define $r_{99,s}$ using the positive radial
diagnostic
\begin{equation}
 \begin{aligned}
  \int_{r_+}^{r_{99,s}}\rho_s(r)\,\dd r
  &=0.99\int_{r_+}^{\infty}\rho_s(r)\,\dd r,\\
  \rho_s(r)&=(r^2+a^2)|R_s(r)|^2.
 \end{aligned}
 \label{eq:r99}
\end{equation}
and take $r_{99}=\max(r_{99,i},r_{99,f})$ for comparison with the orbital
separation $b$.  The displayed tidal region requires $r_{99}/b\le0.10$,
$M\Omega_{\rm res}\le10^{-2}$, and
$|\dot\Omega|/\Omega_{\rm res}^2\ll1$.  The first condition ensures that the
metric expansion covers the main support of the cloud, the second keeps the
orbit in the slowly varying near zone, and the third justifies an adiabatic
tidal field.  Figure~\ref{fig:saturation} shows the saturation branches and
the exact resonance frequencies.  The small values of $M\Omega_{\rm res}$
make the orbital tide adiabatic even when the cloud modes themselves require
a relativistic treatment.

\begin{table*}[t]
\caption{Transition channels on their respective saturation branches. The tidal-domain interval contains values of $\alpha$ for which at least one mass ratio in $10^{-4}\le q\le1$ satisfies $r_{99}/b\le0.10$ and the adiabatic condition. The correction is evaluated relative to the hydrogenic quadrupole matrix element within that interval. Strict samples additionally satisfy $r_{99}/b\le0.07$, retain a correction above $10\%$, and exceed three times the combined numerical and tidal uncertainty.}
\label{tab:channels}
\small
\begin{ruledtabular}
\begin{tabular}{lcccc}
Transition & prograde $\alpha$ & tidal-domain $\alpha$ & correction (\%) & strict samples \\
\hline
$|211\rangle\to|21{-1}\rangle$ & 0.100--0.450 & 0.100--0.225 & $[-15.0,-2.9]$ & 1 \\
$|322\rangle\to|320\rangle$ & 0.100--0.450 & 0.100--0.450 & $[-21.7,-1.1]$ & 3 \\
$|322\rangle\to|300\rangle$ & 0.100--0.350 & --- & --- & 0 \\
$|433\rangle\to|431\rangle$ & 0.100--0.450 & 0.100--0.450 & $[-11.3,-0.6]$ & 2 \\
$|433\rangle\to|411\rangle$ & 0.100--0.450 & 0.100--0.125 & $[-0.5,-0.3]$ & 0 \\
\end{tabular}
\end{ruledtabular}
\end{table*}

\section{Relativistic transition matrix elements}
\label{sec:matrix}

Quasibound modes obey radiative boundary conditions.  Their projection is
therefore based on the conserved bilinear form associated with the discrete
$t$--$\phi$ reflection $\mathcal J$
\cite{Green:2022htq,Cannizzaro:2023jle},
\begin{equation}
 \bbra{\Phi_p}{\Phi_q}
 =\int_{\Sigma}\omega^a(\Phi_p,\mathcal J\Phi_q)\,\dd\Sigma_a,
 \qquad \bbra{\Phi_p}{\Phi_q}=\delta_{pq}.
 \label{eq:bilinear}
\end{equation}
In this bilinear product, the second mode enters unconjugated.  Near-horizon
contributions are defined by analytic continuation of the radial coordinate,
which makes the surface integral finite and preserves the symmetry of the
differential operator.  Appendix~\ref{app:bilinear} gives the contour
prescription and its relation to a finite-part evaluation.

Let $g_{ab}\to g_{ab}+h_{ab}$ and expand the field in normalized Kerr modes,
$\Phi=\sum_j c_j(t)\Phi_j$.  To first order in the tide, slowly varying
amplitudes obey
\begin{equation}
 i\dot c_f=\sum_i\eta_{fi}(t)c_i,
 \qquad
 \eta_{fi}=-\frac{\bbra{\Phi_f}{\delta\Box_g\Phi_i}}
 {\bbra{\Phi_f}{\Phi_f}}.
 \label{eq:amplitude}
\end{equation}
The variation of the scalar wave operator may be written as
\begin{equation}
 \delta\Box_g\Phi=-h^{ab}\nabla_a\nabla_b\Phi
 -\left(\nabla_a h^{ab}-\frac12\nabla^b h\right)\nabla_b\Phi,
 \label{eq:dbox}
\end{equation}
or in the equivalent divergence form given in Appendix~\ref{app:ward}.
Using both forms is useful because their cancellations occur differently in
the radial and angular integrations.

The adiabatic Kerr quadrupole is written
\begin{equation}
 \begin{aligned}
 h_{ab}&=\epsilon_2\left[\widehat h^{(2,-2)}_{ab}(r,\theta)
 e^{-2i(\phi-\Omega v)}+\mathrm{c.c.}\right],\\
 \epsilon_2&=\frac{M^2M_*}{b^3}
 =\frac{q}{1+q}(M\Omega)^2.
 \end{aligned}
 \label{eq:kerrtide}
\end{equation}
where $q=M_*/M$ and the perturbation is in ingoing radiation gauge
\cite{Katagiri:2026}.  Retaining the helical phase in advanced time $v$
keeps the time derivatives in Eq.~\eqref{eq:dbox}.  We define the
dimensionless complex transition matrix element for a unit quadrupolar tide
by
\begin{equation}
 \Kfi=-\int_{\Sigma,\mathcal C}(\mathcal J\Phi_f)
 \delta\Box_{\widehat h}\Phi_i\,\dd\Sigma,
 \qquad M\eta_{fi}=\epsilon_2\Kfi.
 \label{eq:kernel}
\end{equation}
A rephasing of either mode changes the phase of $\Kfi$ and leaves its
magnitude unchanged.  All single-transition comparisons below therefore use
$|\Kfi|$.

The on-resonance matrix element is invariant under smooth gauge changes with
compact support.  For $h_{ab}=\Lie_\xi g_{ab}$, covariance gives
\begin{equation}
 \delta(\Box_g-\mu^2)_{\Lie_\xi g}
 =[\Lie_\xi,\Box_g-\mu^2].
 \label{eq:wardoperator}
\end{equation}
After applying the unperturbed field equation and the bilinear symmetry, the
pure-gauge contribution takes the form
\begin{equation}
 \mathcal K_{fi}[\Lie_\xi g]
 =\bigl[\omega_i-\omega_f-(m_i-m_f)\Omega\bigr]
 \mathcal B_{fi}[\xi].
 \label{eq:ward}
\end{equation}
It vanishes at Eq.~\eqref{eq:resonance}; the boundary functional
$\mathcal B_{fi}$ is finite because $\xi^a$ has compact support.  This
identity also explains why keeping the helical time dependence is essential.
The detailed derivation is in Appendix~\ref{app:ward}.

To isolate physical sources of the correction, we evaluate three nested
matrix elements:
\begin{align}
 \mathcal K_{\rm H}&=\mathcal K[\text{hydrogenic modes, Newtonian tide}],
 \nonumber\\
 \mathcal K_{\rm N}&=\mathcal K[\text{Kerr modes, Newtonian tide}],
 \nonumber\\
 \mathcal K_{\rm rel}&=\mathcal K[\text{Kerr modes, Kerr tide}].
 \label{eq:layers}
\end{align}
The first is the conventional gravitational-atom limit, the second measures
the influence of the exact cloud, and the third adds the relativistic tidal
geometry.  The same exact frequencies and saturation spin are used in the
last two layers.

\section{Relativistic transition spectrum}
\label{sec:spectrum}

\begin{figure*}[t]
 \includegraphics[width=\textwidth]{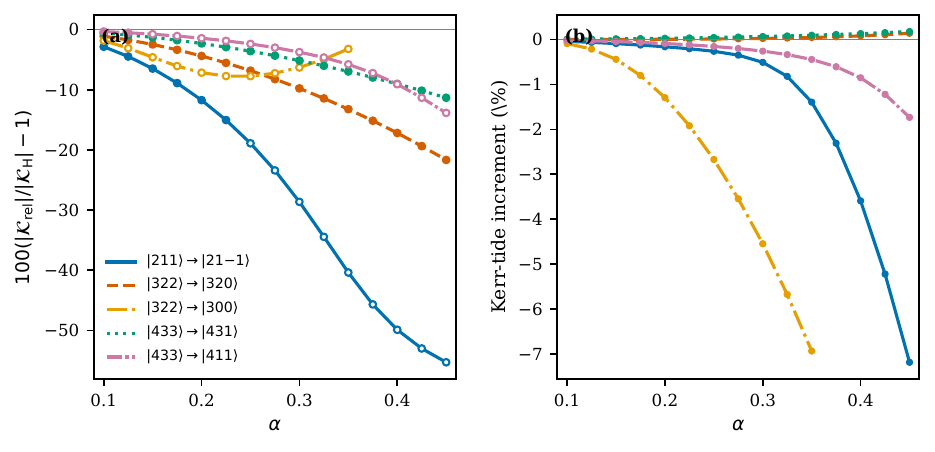}
 \caption{Relativistic transition spectrum.  (a) Fractional change of the
 Kerr-tide matrix element relative to the hydrogenic quadrupole result.
 Filled symbols have at least one mass ratio satisfying $r_{99}/b\le0.10$;
 open symbols show the continuation outside that region.  Numerical error
 bars are included on the filled symbols and are smaller than the marker
 size.  (b)
 Additional change produced by replacing the Newtonian quadrupole with the
 adiabatic Kerr tide while retaining the same Kerr modes.}
 \label{fig:spectrum}
\end{figure*}

Figure~\ref{fig:spectrum} presents the central result.  The survey contains
75 combinations of coupling and transition channel.  Of these,
$\TidalDomainCount$ have an allowed mass ratio for which the spatial and
adiabatic conditions of Sec.~\ref{sec:clouds} hold.  Table~\ref{tab:channels}
summarizes their ranges.  The relativistic result is smaller in magnitude
than the hydrogenic result throughout these ranges.  The suppression grows
smoothly with $\alpha$, reaching $15.0\%$ for $211\to21{-1}$ and
$\MaxKernelCorrection$ for $322\to320$.  The $433\to431$ channel reaches
$11.3\%$, while $433\to411$ remains below one percent in its comparatively
narrow tidal region.

The two strongest channels illustrate complementary aspects of the cloud.
For $211\to21{-1}$, the initial and final states share the same leading
hydrogenic radial function.  Their Newtonian overlap is consequently simple,
but the Kerr radial equation responds differently to $m=1$ and $m=-1$ near
the horizon.  The bilinear projection weights this region with the analytic
continuation appropriate to each mode.  In $322\to320$, the two states again
share $(n,l)$ but have a larger cloud and a more rapidly growing relativistic
splitting across the upper part of the saturation branch.  The accumulated
radial displacement reduces the overlap even though the spheroidal angular
functions remain close to spherical harmonics.

The cross-multiplet channel $322\to300$ provides a useful contrast.  It has a
prograde resonance only through $\alpha=0.35$, and its extended final state
prevents $r_{99}/b$ from becoming small over $10^{-4}\le q\le1$.  Scale
separation explains this behavior:
the small level splitting moves the resonance to a wide cloud-to-orbit ratio.
For $433\to411$, a similar geometric restriction leaves only the two lowest
couplings in the displayed tidal region.  An adiabatic local tide must
therefore satisfy the spatial coverage condition in addition to admitting a
prograde level crossing.

We quantify the origin of the correction through a component-resolved
comparison.  Within the Newtonian tidal operator, the mode frequency, radial
profile, angular profile, and bilinear normalization are replaced one at a
time and in all possible orders.  For a set $S$ of replaced components, let
$v(S)=\ln|\mathcal K_S/\mathcal K_{\rm H}|$.  The order-averaged contribution
of component $a$ is
\begin{equation}
 \begin{aligned}
 \varphi_a={}&\sum_{S\subseteq N\setminus\{a\}}
 \frac{|S|!(|N|-|S|-1)!}{|N|!}\\
 &\times[v(S\cup\{a\})-v(S)],\\
 \sum_a\varphi_a={}&\ln\left|
 \frac{\mathcal K_{\rm N}}{\mathcal K_{\rm H}}\right|.
 \end{aligned}
 \label{eq:components}
\end{equation}
This decomposition compares complete matrix elements, and the component
values serve as bookkeeping contributions.

\begin{figure*}[t]
 \includegraphics[width=\textwidth]{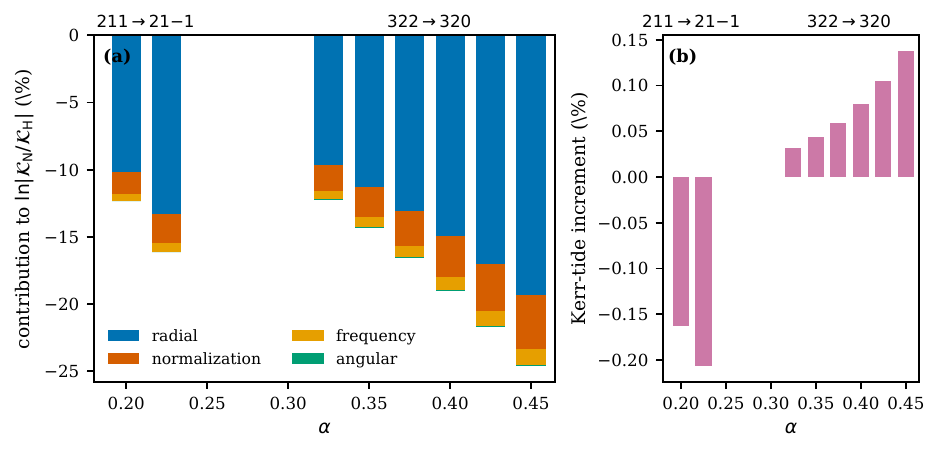}
 \caption{Physical origin of the relativistic correction at eight selected
 samples.  (a) Order-averaged contributions of the radial
 profile, bilinear normalization, mode frequency, and angular profile to the
 logarithmic suppression.  (b) Additional percentage change from the Kerr
 quadrupolar geometry.}
 \label{fig:origin}
\end{figure*}

As shown in Fig.~\ref{fig:origin}, the radial profile accounts for
$\RadialShareRange$ of the logarithmic suppression at the eight points
selected for the component-resolved comparison.  The bilinear normalization
accounts for another
$\NormalizationShareRange$.  Frequency and angular changes are small.  This
hierarchy has a direct spatial interpretation.  Kerr corrections to the
bound-state equation accumulate over the entire radial support of the cloud,
and the bilinear norm changes the global weighting of that profile.  By
contrast, departures of the tidal metric from its Newtonian quadrupole are
largest near the black hole, where only a small fraction of the extended
cloud contributes.  The resulting Kerr-tide increment is
$\GeometryIncrementRange$, as shown in the right panel.

The six points in Table~\ref{tab:strict} retain a suppression above $10\%$
under the tighter condition $r_{99}/b\le0.07$.  They occur in the
$211\to21{-1}$, $322\to320$, and $433\to431$ channels.  Numerical
displacements are many orders of magnitude below the physical correction,
while the exact Newtonian multipole displacement is below $0.1\%$.  These
comparisons show that the effect persists for clouds well inside the tidal
region.

\begin{table*}[t]
\caption{Representative matrix elements that retain a correction above $10\%$ within $r_{99}/b\le0.07$ and exceed three times the combined numerical and tidal uncertainty. The numerical column is the largest relative displacement under one-at-a-time resolution and integration changes. The multipole column is the largest exact-Newtonian displacement from the quadrupole result over the listed mass-ratio interval.}
\label{tab:strict}
\small
\begin{ruledtabular}
\begin{tabular}{lcccccc}
Transition & $\alpha$ & $\chi_{\rm sat}$ & $M\Omega_{\rm res}$ & $q$ interval & $\Delta|\mathcal K|/|\mathcal K_H|$ & numerical / multipole (\%) \\
\hline
$|211\rangle\to|21{-1}\rangle$ & 0.200 & 0.687045 & $4.843e-06$ & $0.49$--$1$ & -11.73\% & 5.98e-07 / 0.000 \\
$|322\rangle\to|320\rangle$ & 0.325 & 0.584995 & $4.031e-06$ & $0.0001$--$1$ & -11.45\% & 2.84e-06 / 0.044 \\
$|322\rangle\to|320\rangle$ & 0.350 & 0.620127 & $6.864e-06$ & $0.0001$--$1$ & -13.24\% & 2.33e-06 / 0.066 \\
$|322\rangle\to|320\rangle$ & 0.375 & 0.653444 & $1.130e-05$ & $0.2$--$1$ & -15.15\% & 1.68e-06 / 0.085 \\
$|433\rangle\to|431\rangle$ & 0.425 & 0.521945 & $2.597e-06$ & $0.0001$--$1$ & -10.16\% & 6.04e-06 / 0.036 \\
$|433\rangle\to|431\rangle$ & 0.450 & 0.547433 & $3.899e-06$ & $0.0001$--$1$ & -11.34\% & 3.14e-06 / 0.049 \\
\end{tabular}
\end{ruledtabular}
\end{table*}

\section{Robustness and binary consequences}
\label{sec:robustness}

\subsection{Spatial expansion and higher multipoles}

\begin{figure*}[t]
 \includegraphics[width=\textwidth]{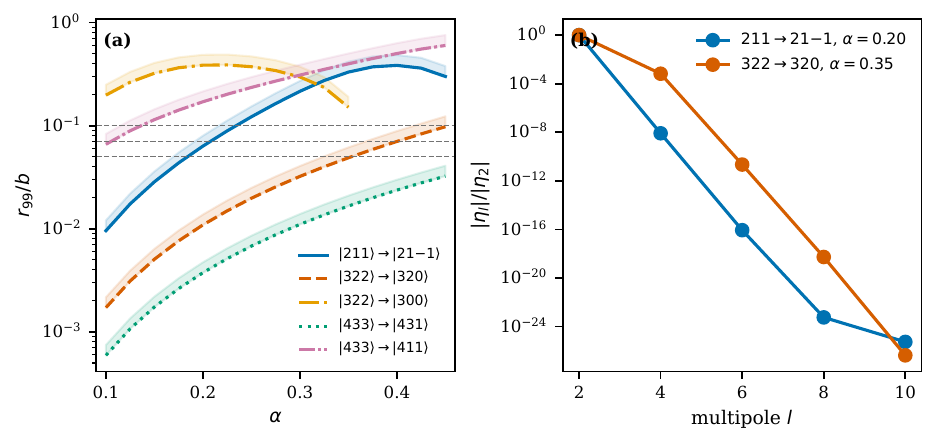}
 \caption{Spatial validity and higher tidal multipoles.  (a) Range of
 $r_{99}/b$ across $10^{-4}\le q\le1$ for each transition.  Horizontal lines
 mark the reference levels $r_{99}/b=0.05$, $0.07$, and $0.10$.  (b)
 Magnitudes of exact
 Newtonian multipoles relative to the quadrupole for two representative
 transitions.  The inner and outer radial branches of the point-mass
 potential are both included.}
 \label{fig:robustness}
\end{figure*}

The ratio $r_{99}/b$ varies with both $\alpha$ and mass ratio.  Panel (a) of
Fig.~\ref{fig:robustness} shows the full range obtained for
$10^{-4}\le q\le1$.  The $322\to320$ resonance benefits from an increasing
level splitting at larger $\alpha$, which reduces $b$ more slowly than the
cloud contracts.  This creates a broad interval in which the relativistic
matrix-element effect and the spatial expansion are simultaneously under
control.  The $211\to21{-1}$ hyperfine splitting is smaller, so its strict
region appears only at larger mass ratios.

As a direct test of the quadrupole approximation, we also use the point-mass
tidal potential after subtracting the monopole and dipole terms,
\begin{equation}
 \begin{aligned}
 V_{\rm tide}&\equiv-\frac{\mu M_*}{|\bm x-\bm b|}-V_0-V_1,\\
 V_{\rm tide}&=-4\pi\mu M_*
 \sum_{l=2}^{\infty}\sum_{m=-l}^{l}\frac{1}{2l+1}\\
 &\times\frac{r_<^l}{r_>^{l+1}}Y_{lm}(\theta,\phi)
 Y^*_{lm}(\tfrac{\pi}{2},\Omega t).
 \end{aligned}
 \label{eq:pointpotential}
\end{equation}
The monopole is a common energy shift, and the dipole is removed in the
freely falling frame centered on the black hole.  The radial integrals are
split at $r=b$, preserving both the interior and exterior branches.
For the $m_*=2$ electric tide, $P_l^2(0)=0$ removes odd $l$ analytically.
Even multipoles through $l=10$ converge rapidly, as seen in panel (b).  At
$\alpha=0.35$ for $322\to320$, their net displacement from the quadrupole is
$\MultipoleRepresentativeShift$.  This is two orders of magnitude smaller
than the relativistic cloud correction.

The covariant calculation is retained at quadrupolar order.  The exact
Newtonian sum bounds the sensitivity to additional spatial structure of the
companion field in the region where $M/b$ is small.  A conservative estimate
twice the exact Newtonian shift remains well below the $10\%$ effects in
Table~\ref{tab:strict}.  Strong-field corrections to higher multipoles enter
with the same additional powers of $M/b$ and remain subdominant to the leading
Kerr-mode contribution for these points.

\subsection{Independent radial construction}

The quasibound-state normalization is the least familiar ingredient in the
calculation, so it is checked by a second radial construction.  The complex
frequencies are held fixed, while angular functions are obtained from a
boundary-value spheroidal equation and radial functions are shot inward from
the quasibound asymptotic expansion and outward from the horizon Frobenius
expansion.  The solutions are matched at several decay lengths.  Their norm
and matrix element are then integrated along deformed complex contours,
without using the finite-part radial representation of the central
calculation.  Across the five Kerr transitions, the maximum relative
amplitude difference is $\IndependentKernelDifference$, and the phase
differences are below $3\times10^{-9}\,\mathrm{rad}$.  Contour deformation changes the result by less than
$2\times10^{-3}$, and the matched Wronskian is below $10^{-8}$.

In the Schwarzschild limit, the $211\to31{-1}$ comparison reproduces the
relativistic correction reported in Ref.~\cite{Cannizzaro:2023jle} within
the digitization uncertainty.  This control probes a different radial
overlap and connects the present normalization directly to the published
semirelativistic result.  Appendix~\ref{app:comparisons} collects these
comparisons and the resolution dependence of the frequencies, eigenfunctions,
and matrix elements.

\subsection{Landau--Zener depletion}

To illustrate the binary consequence of the corrected matrix elements, we
prescribe a circular gravitational-wave inspiral.  Kepler's relation and the
leading chirp are
\begin{equation}
 \begin{aligned}
 \frac{b}{M}&=(1+q)^{1/3}(M\Omega)^{-2/3},\\
 M^2\dot\Omega_{\rm N}&=\frac{96}{5}
 \frac{q}{(1+q)^{1/3}}(M\Omega)^{11/3}.
 \end{aligned}
 \label{eq:chirp}
\end{equation}
The two-level Landau--Zener parameter and depletion probability are
\begin{equation}
 z=\frac{|M\eta_{fi}|^2}{|\Delta m|M^2\dot\Omega},
 \qquad P_{\rm dep}=1-e^{-2\pi z}.
 \label{eq:lz}
\end{equation}
These formulae follow the gravitational-atom treatment
\cite{Landau1932,Zener1932,Baumann:2019ztm}.  We also vary the sweep rate by
the standard 1PN circular correction \cite{Blanchet:2013haa}, given explicitly
in Appendix~\ref{app:lz}.

\begin{figure*}[t]
 \includegraphics[width=\textwidth]{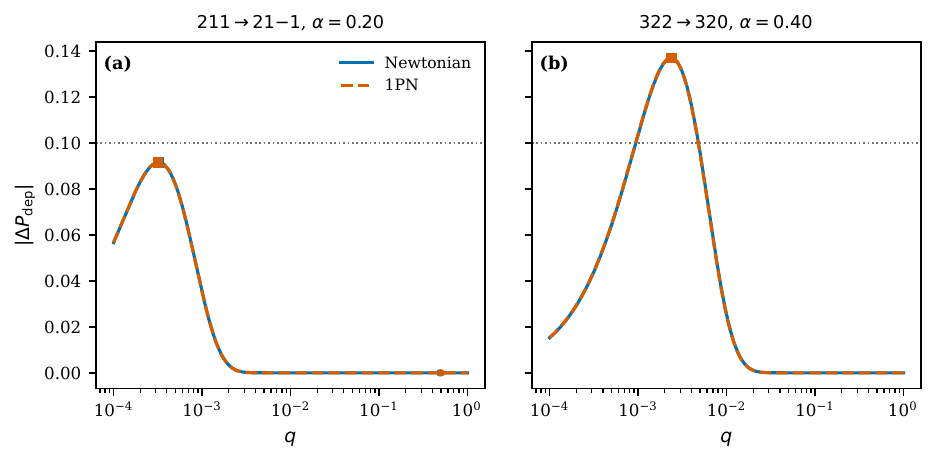}
 \caption{Absolute change in Landau--Zener depletion produced by the
 relativistic matrix element.  Filled circles lie within $r_{99}/b\le0.07$;
 squares mark the maximum along each curve.  Solid and dashed lines use the
 Newtonian and 1PN sweep rates, respectively.  The horizontal line denotes a
 $10\%$ change.}
 \label{fig:depletion}
\end{figure*}

At fixed orbit, $z\propto|\mathcal K|^2$.  The response to a small
matrix-element shift is therefore
\begin{equation}
 \delta\ln z=2\,\delta\ln|\mathcal K|,
 \qquad
 \delta P_{\rm dep}=4\pi z e^{-2\pi z}
 \delta\ln|\mathcal K|.
 \label{eq:sensitivity}
\end{equation}
This relation explains the shape of Fig.~\ref{fig:depletion}.  In the
diabatic limit the absolute depletion is small; in the adiabatic limit both
predictions saturate near unity.  The largest observable difference occurs
for $z$ of order unity.  The $211\to21{-1}$ example reaches a $9.2\%$
difference, while the $\alpha=0.40$ $322\to320$ resonance reaches
$\MaxDepletionChange$.  The Newtonian and 1PN sweep rates give nearly the
same maximum, indicating that the matrix-element correction controls the
effect while the 1PN change in sweep rate has little influence.

The depletion curves are a local use of the transition spectrum within a
prescribed inspiral.  In a complete binary history, repeated resonances,
decaying-mode absorption, cloud backreaction, and changes in black-hole spin
can alter the occupation arriving at a given crossing
\cite{Berti:2019wnn,Tong:2022bbl,Takahashi:2023,Tomaselli:2024bdd,
Tomaselli:2025wide,Boskovic:2025trails}.  In comparable-mass systems,
stellar common-envelope evolution can help a cloud survive to smaller
separations, where cloud mass transfer or a cloud common-envelope phase may
also become relevant \cite{Guo:2024iye,Guo:2025ckp}.  These effects operate
on the amplitudes after the microscopic matrix element has been supplied.
The spectrum calculated here can therefore be inserted directly into such
evolution equations.

\section{Discussion and conclusions}
\label{sec:discussion}

Tidal spectroscopy of a saturated boson cloud is sensitive to the
relativistic structure of the cloud well before the companion probes the
near-horizon geometry directly.  Across the transitions studied here, the
dominant departure from the hydrogenic estimate comes from the radial Kerr
quasibound functions.  Their global deformation changes the overlap over the
full cloud, and the bilinear norm supplies a smaller coherent correction.
The angular spheroidal deformation and the relativistic quadrupolar geometry
are comparatively weak.  This hierarchy explains why an accurate cloud
basis matters even when $r_{99}/b$ is small enough for an adiabatic tidal
expansion.

The result sharpens the microscopic input needed by binary-cloud evolution
models.  A ten-percent change in a transition amplitude produces a
twenty-percent change in the Landau--Zener exponent, but the corresponding
occupation change depends on where the resonance lies between the diabatic
and saturated limits.  The examples here reach an absolute depletion change
of order ten percent.  Such changes can alter which cloud survives to later
resonances and hence affect the sequence of angular-momentum exchanges that
underlies floating, sinking, or resonantly modified inspirals
\cite{Cardoso:2011xi,Ferreira:2017pth,Zhang:2019eid}.  Environmental effects
of clouds are also relevant to extreme-mass-ratio signals
\cite{Cardoso:2022whc,Brito:2023pyl,Duque:2023seg,Dyson:2025dlj,Li:2025ffh}
and to population-level searches with terrestrial and space detectors
\cite{DellaMonica:2025zby,Audley:2017drz,Babak:2017tow}.

The local adiabatic tide resolves the portion of the metric sampled by a
compact cloud and exposes the strong-field wave-function effect cleanly.
The scale tests in Sec.~\ref{sec:robustness} show that higher Newtonian
multipoles are too small to account for the observed suppression.  A global
binary perturbation would refine the spatial transition between the local
Kerr geometry and the companion zone, while a self-consistent orbit would
determine the occupation and sweep rate at each crossing.  Accretion and
other environmental torques can likewise modify the long inspiral
\cite{Kocsis:2011dr,Yunes:2011ws,Speri:2022upm,Cole:2022yzw}.  These additions
primarily change the binary trajectory through the transition spectrum;
the Kerr quasibound matrix elements remain its local field-theory input.

The present calculation establishes a relativistic spectrum for five
transitions on three saturated-cloud branches.  It identifies a broad
$322\to320$ region and a smaller $211\to21{-1}$ region where the hydrogenic
amplitude misses a physically robust effect above ten percent.  Exact
companion multipoles and an independent complex-contour construction support
the result.  Relativistic Kerr cloud modes should consequently replace
hydrogenic wave functions in precision calculations of resonant cloud
depletion and binary evolution.

\begin{acknowledgments}
This work is supported by the Tianshan Talent Training Program (Grant
No.~2023TSYCCX0099) and the National Key R\&D Program of China (Grant
No.~2024YFA1611503).  This work is also partly supported by the Urumqi Nanshan
Astronomy and Deep Space Exploration Observation and Research Station of
Xinjiang (XJYWZ2303) and the Central Guidance for Local Science and Technology
Development Fund (Grant No.~ZYYD2026JD01).
\end{acknowledgments}

\section*{Data Availability}

The code and numerical data supporting this article are available through the
public repository and archived release in Ref.~\cite{Li2026Data}.  The cited
Zenodo record identifies the exact software and data version used for the
results presented here.

\appendix

\section{Kerr quasibound states}
\label{app:qbs}

The massive scalar obeys $(\Box_g-\mu^2)\Phi=0$.  With the convention in
Eq.~\eqref{eq:mode}, the separated angular and radial equations are
\begin{align}
 &\frac{1}{\sin\theta}\frac{\dd}{\dd\theta}
 \left(\sin\theta\frac{\dd S}{\dd\theta}\right)
 +\left[c^2\cos^2\theta-\frac{m^2}{\sin^2\theta}+A_{lm}\right]S=0,
 \label{eq:angular}\\
 &\frac{\dd}{\dd r}\left(\Delta\frac{\dd R}{\dd r}\right)
 +\left[\frac{\omega^2(r^2+a^2)^2}{\Delta}
 -\frac{4Mam\omega r-m^2a^2}{\Delta}\right.\nonumber\\
 &\hspace{4.0cm}\left.-a^2\omega^2-\mu^2r^2-A_{lm}\right]R=0,
 \label{eq:radial}
\end{align}
where $\Delta=r^2-2Mr+a^2$.  These equations follow from separability of
Kerr perturbations \cite{Teukolsky:1972my,Teukolsky:1973ha}.  With
$k=(\mu^2-\omega^2)^{1/2}$, $\mathrm{Re}\,k>0$,
$\kappa_H=(r_+-r_-)/[2(r_+^2+a^2)]$, and
$\nu_\infty=M(2\omega^2-\mu^2)/k$, the quasibound boundary conditions are
\begin{equation}
 \begin{aligned}
 R&\sim(r-r_+)^{-i(\omega-m\Omega_H)/(2\kappa_H)}
 &&(r\to r_+),\\
 R&\sim r^{\nu_\infty-1}e^{-kr}&&(r\to\infty).
 \end{aligned}
 \label{eq:qbsbc}
\end{equation}

Mapping the radial coordinate to $x=(r-r_+)/(r-r_-)$ and factoring these
asymptotic powers gives a series $R=e^{-kr}x^\rho(r-r_-)^\sigma
\sum_{j=0}^\infty a_jx^j$.  Its coefficients obey
\begin{equation}
 \alpha_j a_{j+1}+\beta_j a_j+\gamma_j a_{j-1}=0,
 \qquad
 0=\beta_0-\frac{\alpha_0\gamma_1}{\beta_1-}
 \frac{\alpha_1\gamma_2}{\beta_2-}\cdots .
 \label{eq:cf}
\end{equation}
The coefficients are those of the massive scalar recurrence in
Ref.~\cite{Dolan:2007mj}.  We evaluate the fraction both downward from a large
index and through a modified continued-fraction iteration.  Frequencies are
continued in $\alpha$ and $\chi$, and Eq.~\eqref{eq:saturation} is then
solved along each $(n,l,m)$ branch.  Increasing the series order leaves the
real frequency stable below $10^{-8}$ relative and the resolvable imaginary
part below $10^{-3}$ relative.

The hydrogenic limit provides an analytic check,
\begin{equation}
 M\omega_{nlm}=\alpha\left(1-\frac{\alpha^2}{2n^2}
 +O(\alpha^4)\right)+iM\Gamma_{nlm}.
 \label{eq:hydrogenic}
\end{equation}
It recovers the spherical eigenvalue $A_{lm}\to l(l+1)$ and the expected
Bohr radii as $\alpha\to0$.  The fully relativistic frequencies are used for
all saturation and resonance conditions in the main text.

\section{Bilinear projection and contour normalization}
\label{app:bilinear}

For two Klein--Gordon solutions, the symplectic current is
\begin{equation}
 \omega^a(\Phi_1,\Phi_2)=
 \Phi_1\nabla^a\Phi_2-\Phi_2\nabla^a\Phi_1.
 \label{eq:symplectic}
\end{equation}
Combining this current with the $t$--$\phi$ reflection maps the outgoing
adjoint problem back to the original Kerr problem.  The resulting bilinear
form, Eq.~\eqref{eq:bilinear}, is symmetric and renders nondegenerate modes
orthogonal.  No complex conjugation is introduced in the radial product.

The horizon behavior makes a constant-real-$r$ integral divergent for a
generic complex frequency.  Analytic continuation in the tortoise coordinate
chooses a contour $\mathcal C$ satisfying
\begin{equation}
 \arg r_*+\arg(\omega_p+\omega_q)=-\frac{\pi}{2},
 \qquad r_*\to-\infty,
 \label{eq:contour}
\end{equation}
so the continued integrand decays.  Deforming $\mathcal C$ within an analytic
sector leaves the result unchanged.  Equivalently, one may subtract the
near-horizon Frobenius divergences, integrate the remainder on real $r$, and
add their analytic finite parts.  Agreement of these two constructions
tests both the normalization and the radial continuation.

Projecting the perturbed field equation and neglecting second derivatives of
the slowly varying amplitudes gives Eq.~\eqref{eq:amplitude}.  The normalization
in Eq.~\eqref{eq:bilinear} removes any arbitrary scaling of $R$ or $S$.
An independent rephasing $\Phi_j\to e^{i\vartheta_j}\Phi_j$ sends
$\mathcal K_{fi}\to e^{i(\vartheta_i-\vartheta_f)}\mathcal K_{fi}$, making
$|\mathcal K_{fi}|$ the invariant quantity for a single link.

\section{Wave-operator variation and the Ward identity}
\label{app:ward}

Varying $g^{ab}$ and $\sqrt{-g}$ in
$\Box_g\Phi=(-g)^{-1/2}\partial_a(\sqrt{-g}g^{ab}\partial_b\Phi)$ yields
\begin{equation}
 \delta\Box_g\Phi=-\frac{h}{2}\Box_g\Phi
 +\frac{1}{\sqrt{-g}}\partial_a\left\{\sqrt{-g}
 \left[\frac{h}{2}g^{ab}-h^{ab}\right]\partial_b\Phi\right\}.
 \label{eq:dboxdiv}
\end{equation}
Expanding the derivative and using metric compatibility gives
Eq.~\eqref{eq:dbox}.  This algebraic equivalence holds before the field
equations are imposed.

For a diffeomorphism generated by $\xi^a$, scalar covariance implies
\begin{equation}
 \begin{aligned}
 &(\Box_{g+\Lie_\xi g}-\mu^2)(\Phi+\Lie_\xi\Phi)\\
 &\quad=(\Box_g-\mu^2)\Phi
 +\Lie_\xi[(\Box_g-\mu^2)\Phi]+O(\xi^2),
 \end{aligned}
 \label{eq:covariance}
\end{equation}
which rearranges to Eq.~\eqref{eq:wardoperator}.  Insert two separated modes,
integrate by parts in the bilinear product, and use the harmonic dependence
of a gauge vector proportional to $e^{-im_*\Omega t+im_*\phi}$.  The time and
azimuthal derivatives combine into
$\omega_i-\omega_f-m_*\Omega$, producing Eq.~\eqref{eq:ward}.  Compact
support removes both horizon and outer boundary terms.  Away from resonance
the prefactor is finite, while at resonance the matrix element is unchanged
by the gauge transformation.

\section{Selection rules, exact multipoles, and the weak-field limit}
\label{app:multipoles}

The azimuthal integral gives
\begin{equation}
 \int_0^{2\pi}e^{i(m_i-m_f-m_*)\phi}\,\dd\phi
 =2\pi\delta_{m_*,m_i-m_f}.
 \label{eq:azimuthal}
\end{equation}
The Gaunt coefficient further requires the triangle condition and even
$l_i+l_f+l_*$.  For the equatorial $m_*=2$ harmonic,
$Y_{l2}(\pi/2,0)\propto P_l^2(0)$, and parity gives $P_l^2(0)=0$ for odd
$l$.  The first nonvanishing electric term is consequently $l=2$.

For each even multipole, the radial contribution retains both sides of the
companion orbit,
\begin{equation}
 \begin{aligned}
 M\eta_l&=\epsilon_2\left(\frac{M}{b}\right)^{l-2}K_l,\\
 K_l&\propto\int_{r_+}^{b}r^{l+2}R_fR_i\,\dd r\\
 &\quad+b^{2l+1}\int_b^\infty r^{1-l}R_fR_i\,\dd r.
 \end{aligned}
 \label{eq:multipole}
\end{equation}
The explicit factor $(M/b)^{l-2}$ accounts for the rapid hierarchy in
Fig.~\ref{fig:robustness}.  At small $\alpha$, spheroidal harmonics approach
spherical harmonics, Kerr radial functions approach hydrogenic functions,
and the bilinear normalization approaches the ordinary bound-state
normalization.  A fit of $|\mathcal K_{\rm rel}|/|\mathcal K_{\rm H}|$ to
$c_0+c_2\alpha^2+c_4\alpha^4$ gives $|c_0-1|<10^{-3}$, establishing a common
weak-field limit for the three layers in Eq.~\eqref{eq:layers}.

\section{Numerical comparisons and uncertainty}
\label{app:comparisons}

Mode frequencies are evaluated with successively longer continued fractions,
and the radial and angular functions are compared at increasing collocation
order.  The transition integral is varied independently in radial and
angular resolution, horizon cutoff, near-horizon subtraction order, outer
range, and quadrature rule.  The quoted error bar is the largest absolute
displacement among these one-at-a-time changes; the root-sum-square value is
retained only as a secondary diagnostic.  All matrix elements displayed in
the tidal region have numerical errors far below their relativistic shift.

The independent construction described in Sec.~\ref{sec:robustness} uses
12-term horizon and infinity expansions, matching radii at 8, 12, and 16
decay lengths, and three contour deformations.  Its five Kerr matrix elements
agree with the central values within $5\times10^{-3}$ in magnitude and phase;
the observed differences are substantially smaller.  The exact Newtonian
multipole sum through $l=10$ changes by less than $10^{-4}$ between its last
two truncations.  Direct Gaussian and adaptive integrations agree within
$10^{-4}$.

For each prominent point, the uncertainty attached to the physical
conclusion is the sum of the numerical envelope, the exact-Newtonian
multipole displacement, and twice that displacement as a conservative scale
for uncomputed relativistic higher multipoles.  The six entries of
Table~\ref{tab:strict} retain a matrix-element correction above $10\%$ and
above three times this combined scale.  The separation of numerical and
tidal uncertainties is maintained because they represent different physical
questions.

\section{Inspiral and Landau--Zener formulae}
\label{app:lz}

For a circular binary, define $\nu_{\rm bin}=q/(1+q)^2$ and
$x=[(1+q)M\Omega]^{2/3}$.  The 1PN sweep rate used in Fig.~\ref{fig:depletion}
is
\begin{equation}
 M^2\dot\Omega_{\rm 1PN}=M^2\dot\Omega_{\rm N}
 \left[1-\left(\frac{743}{336}+\frac{11}{4}\nu_{\rm bin}\right)x\right].
 \label{eq:chirp1pn}
\end{equation}
The two-level Hamiltonian near resonance has detuning
$\delta(t)=\Delta\omega-\Delta m\,\Omega(t)$ and off-diagonal element
$\eta_{fi}$.  Linearizing $\delta$ at the crossing gives
$|\dot\delta|=|\Delta m|\dot\Omega$ and hence Eq.~\eqref{eq:lz}.  The
black-hole mass cancels from $z$: changing $M$ rescales physical frequencies
by $M^{-1}$ and times by $M$, while the dimensionless transition spectrum and
depletion curves remain unchanged.

\bibliography{references}

\end{document}